\documentclass[%
aps,reprint,pre,
amsmath,
amssymb,
floatfix,
]{revtex4-2}

\usepackage{float}
\usepackage{graphicx}
\usepackage{dcolumn}
\usepackage{bm}
\usepackage{color}

\newcommand{\mrm}[1]{\mathrm{#1}}
\newcommand{\bbm}[1]{\boldsymbol{\mrm{#1}}}
\newcommand{\er}[0]{ \bbm{\hat{e}}_r }
\newcommand{\et}[0]{ \bbm{\hat{e}}_\theta }
\newcommand{\un}[0]{ \bbm{\hat{n}}}
\newcommand{\nbr}[0]{\mathrm{N_{Br}}}
\newcommand{\nba}[0]{\mathrm{N_{Ba}}}
\newcommand{\sk}[0]{Sk}
\newcommand{\as}[0]{As}
\DeclareMathOperator{\sgn}{sgn}
\DeclareMathOperator{\sech}{sech}

\allowdisplaybreaks

\usepackage[colorlinks=true,breaklinks=true]{hyperref}

\begin{document}

\title{Tuning a magnetic field to generate spinning ferrofluid droplets with controllable speed via nonlinear periodic interfacial waves}

\author{Zongxin Yu}
\email{yu754@purdue.edu}

\author{Ivan C.\ Christov}
\thanks{Corresponding author.}
\email{christov@purdue.edu}
\homepage{http://tmnt-lab.org}

\affiliation{School of Mechanical Engineering, Purdue University, West Lafayette, Indiana 47907, USA}

\date{\today}

\begin{abstract}
Two dimensional free surface flows in Hele-Shaw configurations are a fertile ground for exploring nonlinear physics. Since Saffman and Taylor's work on linear instability of fluid--fluid interfaces, significant effort has been expended to determining the physics and forcing that set the linear growth rate. However, linear stability does not always imply nonlinear stability. We demonstrate how the combination of a radial and an azimuthal external magnetic field can manipulate the interfacial shape of a linearly unstable ferrofluid droplet in a Hele-Shaw configuration. We show that weakly nonlinear theory can be used to tune the initial unstable growth. Then, nonlinearity arrests the instability, and leads to a permanent deformed droplet shape. Specifically, we show that the deformed droplet can be set into motion with a predictable rotation speed, demonstrating nonlinear traveling waves on the fluid-fluid interface. The most linearly unstable wavenumber and the combined strength of the applied external magnetic fields determine the traveling wave shape, which can be asymmetric.
\end{abstract}

\maketitle

\section{Introduction} 
Recently, there has been significant interest in the physics of active and responsive fluids \cite{Mar18,S18}. For example, swimming bacteria can take a suspension of microscopic gears out of equilibrium and extract rectified (useful) work out of an otherwise random system \cite{SAGA10}. One promising approach to creating active fluids with controllable properties and behaviors is by suspending many mechanical microswimmers made from shape-programmable materials \cite{Lum16} and actuating them with an external magnetic field \cite{XZSOD19,GKWLS20}. This actuation mechanism is particularly enticing for biological applications due to the safe operation of magnetic fields in the medical setting (for, \textit{e.g.}, targeted therapies and drug delivery \textit{in vivo}) \cite{NKA10}. Even simpler than a suspension of magnetically-responsive mechanical microswimmers is a suspension of ferrofluid droplets, which can also respond to an external magnetic field \cite{R13_Ferrohydrodynamics,BCM10}. Ferrofluids are colloidal dispersion of ferromagnetic nanoparticles in a carrier liquid, such as water, which can be immiscible when placed in another liquid. However, the ferrofluid droplet's interface motion and response to different types of external magnetic fields is not well understood. Previous work has addressed the linear stability of such fluid--fluid interfaces \cite{RZS83,JGC94}, including stationary shapes \cite{ALM18}, but not a droplet's nonlinear dynamics or controllable motion. Guided by the well-established ability of nonlinearity to ``arrest'' long-wave instabilities \cite{BP98}, we demonstrate, using theory and nonlinear simulation, that it is possible to ``grow'' linearly unstable ferrofluid interfaces into well-defined \emph{permanent} shapes. These permanent shapes, which cannot be further deformed without changing the forcing of the system, can then be considers as \emph{solitary waves}, in the sense of a ``localized wave that propagates along one space direction only, with undeformed shape'' \cite[p.~11]{Re99}. Importantly, unlike previous work discussing traveling waves on a ferrofluid interface in a Cartesian configuration \cite{LM12}, we analyze the \emph{fully nonlinear} dynamics of these waves in a novel configuration, and thus ensure they satisfy the solitary wave definition above. Specifically, we show that the resulting coherent droplet shapes are reproducible and controllable via an external magnetic field. These droplets can be set into rotational motion with velocities predictable by the proposed theory, leading to the possibility of an externally-actuated active fluid suspension.

\section{Governing equations}
\begin{figure}
 \centering
 \includegraphics[keepaspectratio=true,width=\columnwidth]{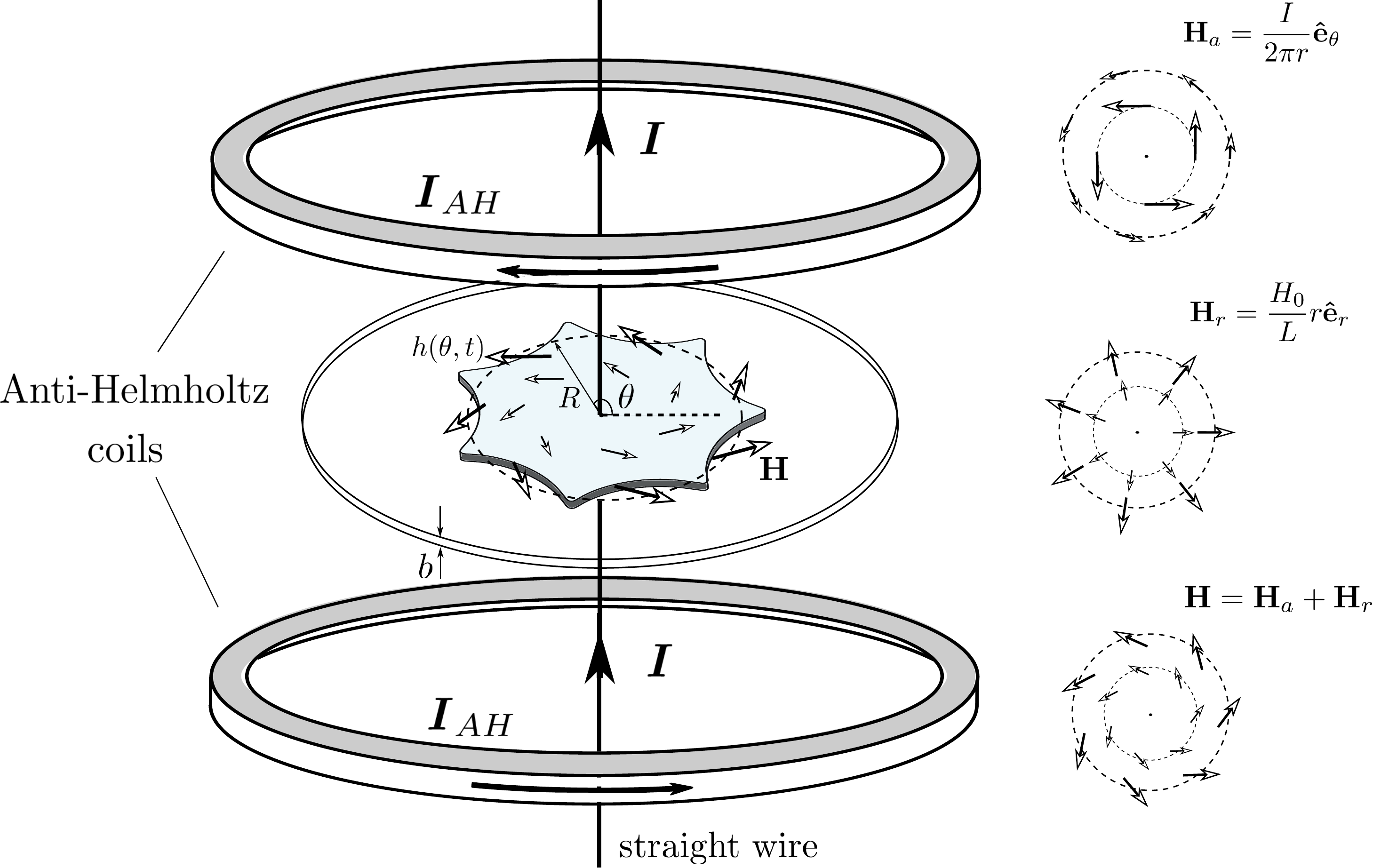}
 \caption{Schematic illustration of a Hele-Shaw cell confining a  ferrofluid droplet, initially circular with radius $R$. The azimuthal magnetic field $\bbm H_{a}$ is produced by a long wire conveying an electric current $I$. The radial magnetic field $\bbm H_{r}$ is produced by a pair of anti-Helmholtz coils with equal currents $I_{AH}$ in opposite directions. The combined external magnetic field $\bbm H$ deforms the droplet, and its interface is given by  $h(\theta,t)$. In comparison, the fluid exterior to the droplet (\textit{e.g.}, air) is assumed to have negligible viscosity and velocity.}
\label{fig:config}
\end{figure}

We study the dynamics of an initially circular ferrofluid droplet (radius $R$) confined in a Hele-Shaw cell with gap thickness $b$ and surrounded by air (negligible viscosity), as shown in Fig.~\ref{fig:config}, because ``[i]f any [ferro]fluid mechanics problem is likely to be accessible to theory and to direct comparison of theory and experiment it should be this one'' \cite{BKLST86}. Both fluids are considered incompressible. We propose to apply the radially-varying external magnetic field 
\begin{equation}
    \bbm H = \underbrace{\frac{I}{2\pi r} \,\et}_{\bbm H_a} + \underbrace{\frac{H_0}{L}r \,\er}_{\bbm H_r}.
\end{equation}
A long wire through the origin, carrying an electric current $I$, produces the azimuthal component $\bbm H_a$. Anti-Helmholtz coils produce the radial component $\bbm H_r$, where $H_0$ is a constant and $L$ is a length scale~\cite{LM16,ALM18}. The combined magnetic field $\bbm H =\bbm H_a+\bbm H_r$ forms an angle with the initially undisturbed interface~\cite{MO04}. The droplet experiences a body force $\propto |\bbm M| \bbm\nabla |\bbm H|$, where $\bbm M$ is the magnetization. To study shape dynamics, we assume the ferrofluid is uniformly magnetized, $\bbm M=\chi \bbm H$, where $\chi$ is its constant magnetic susceptibility. So, $\bbm\nabla |\bbm H|\ne \bbm 0$ is the main contribution to the body force, and the demagnetizing field is negligible, as shown in previous work \cite{MO04,RE06,LM16,ACLM19}.

Enforcing no-slip on the confining boundaries and neglecting inertial terms, the confined flow is governed by a modified Darcy's law~\cite{LM16} with gap-averaged velocity:
\begin{equation}
\bbm v = -\frac{b^2}{12\eta} \bbm\nabla\left(p-\Psi\right), \qquad
\bbm\nabla \cdot \bbm v = 0,
\label{eq:governing}
\end{equation}
where $p$ is the pressure in the droplet, $\eta$ is the ferrofluid's viscosity, $\Psi=\mu_0 \chi |\bbm H|^2/2$ is a scalar potential accounting for the magnetic body force, and $\mu_0$ is the free-space permeability. Here, $\bbm v$ is the velocity field of the ``inner'' ferrofluid, while the viscosity of the ``outer'' fluid is considered negligible (\textit{i.e.}, it is considered inviscid), so the flow exterior to the droplet is neglected. The resulting model is thus, essentially, a one-phase model. 

At the boundary of the droplet, the pressure is given by a modified Young--Laplace law~\cite{R13_Ferrohydrodynamics,BCM10}:
\begin{equation}
p=\tau \kappa-\frac{\mu_0}{2}(\bbm M \cdot \un)^2,
\label{eq:pressure jump}
\end{equation}
where $\tau$ is the constant surface tension, and $\kappa$ is the  curvature of the droplet shape. The second term on the right-hand side of Eq.~\eqref{eq:pressure jump} is the magnetic normal traction~\cite{R13_Ferrohydrodynamics,BCM10}, where $\un$ denotes the outward unit normal vector at the interface. This contribution breaks the symmetry of the initial droplet interface, due to the projection of $\bbm M$ onto $\un$, and causes the droplet to rotate. The kinematic boundary condition 
\begin{equation}
    v_n= -\frac{b^2}{12\eta} \bbm\nabla \left( p-\Psi\right) \cdot \un
    \label{eq:kinematic_bc}
\end{equation} 
requires that the droplet boundary is a material surface.

\section{Mathematical analysis}
We employ the weakly nonlinear approach \cite{MW98} previously adapted to ferrofluid interfacial dynamics (\textit{e.g.}, \cite{MO04,LM16,ALM18}).  The droplet interface is written as $h(\theta,t)=R+\xi(\theta,t)$, where 
\begin{equation}
    \xi(\theta,t)=\sum_{k=-\infty}^{+\infty}\xi_{k}(t)e^{ik\theta}
    \label{eq:xi_def}
\end{equation}
represents the perturbation of the initially circular interface, with complex Fourier amplitudes $\xi_k(t)\in\mathbb{C}$ and azimuthal wavenumbers $k\in\mathbb{Z}$. The velocity potential $\phi=p-\Psi$ is then expanded into a Fourier series as \begin{equation}
    \phi(r,\theta,t)=\sum_{k \neq 0} \phi_k(t)\left(\frac{r}{R}\right)^{|k|}e^{ik\theta},
    \label{eq:phi_Fourier}
\end{equation}
and $\phi_k$ is expressed in terms of $\xi_k$ through the kinematic boundary condition~\eqref{eq:kinematic_bc}. Substituting Eqs.~\eqref{eq:xi_def}, \eqref{eq:phi_Fourier} and \eqref{eq:pressure jump} into Eq.~\eqref{eq:governing}, keeping only terms up to second order in $\xi$, we find the dimensionless equations of motion ($k \neq 0$):
\begin{equation}
\dot{\xi}_k=\Lambda(k)\xi_k 
+\sum_{k' \neq 0}  F(k,k')\xi_{k'} \xi_{k-k'} 
+ G(k,k')\dot{\xi}_{k'} \xi_{k-k'}.
\label{eq:wnl}
\end{equation}
The mode-coupling functions in Eq.~\eqref{eq:wnl} are given by
\begin{subequations}\begin{align}
F(k,k') &= \frac{|k|}{R} \left\lbrace
\frac{\nba}{R^4}[3-\chi k'(k-k')] \right. \nonumber\\
&\qquad + \nbr \lbrace 1+\chi[k'(k-k')+1] \rbrace  
\nonumber\\
& - \left. \frac{1}{R^3}\left[1-\frac{k'}{2}(3k'+k)\right]
+\frac{2\chi \sqrt{\nba \nbr}}{R^2} i k'
\right\rbrace,\\
G(k,k') &= \frac{1}{R}[(\sgn(kk')-1)|k|-1],
\end{align}\end{subequations}%
where $\sgn(x) = x/|x|$ for $x\ne0$ and $\sgn(0)=0$.

From mass conservation, $\xi_0=-\sum_{k> 0}|\xi_{k}|^2/R$ $\forall t\ge0$. Here, 
\begin{multline}
\Lambda(k) = \underbrace{\frac{|k|}{R^3}(1-k^2)}_{\text{surface tension}} - \frac{2\nba }{R^4}|k|
+2(1+\chi)\nbr|k|\\
-\frac{2\chi \sqrt{\nba \nbr}}{R^2}i k |k|
\label{eq:Linear growth}
\end{multline}
denotes the (complex) linear growth rate, and
\begin{equation}
\nba =\frac{\mu_0 \chi I^2}{8 \pi^2 \tau L },\qquad
\nbr =\frac{\mu_0 \chi H_0^2L}{2 \tau }
\end{equation} 
are the magnetic Bond numbers quantifying the ratio of azimuthal and radial magnetic forces to the capillary force, respectively. Terms multiplied by $\chi$ arise from the magnetic normal stress. The time and length scales used in the nondimensionalization are $12\eta L^3/ \tau b^2$ and $L$, respectively.

\section{Linear regime}
First, consider Eq.~\eqref{eq:wnl}, neglecting quadratic terms in $\xi_k$, then  $\Re[\Lambda(k)]=\lambda(k)$ governs the exponential growth or decay of infinitesimal perturbations. For $\lambda(k)>0$, the interface is unstable. Specifically, Eq.~\eqref{eq:Linear growth} indicates that the radial magnetic field term $\propto (1+\chi)\nbr$ is destabilizing, while the azimuthal term $\propto\nba$ and surface tension are stabilizing. The most unstable mode $k_m$ solves $d \lambda(k)/dk=0$:
\begin{equation}
k_m=\sqrt{\frac{1}{3}\left[1-\frac{2\nba}{R}+2(1+\chi)\nbr R^3
\right]}.
\label{eq:km}
\end{equation}
This wavenumber characterizes the dominant $\lfloor k_m \rfloor$-fold symmetry of a pattern. Note that the normal stress from the azimuthal magnetic field does not contribute to the linear dynamics.

The phase velocity of each mode, \begin{equation}
    v_p=-\Im[\Lambda(k)]/k=2\chi \sqrt{\nba \nbr}k/R^2
    \label{eq:linear_vp}
\end{equation}
in the linear regime, is set by $\Im[\Lambda(k)]$. A periodic shape on $[0, 2\pi]$ forms a closed curve, meaning wave propagation is manifested as rotation of the droplet. Motion is caused by the magnetic normal stresses arising from the combined magnetic field. Intuitively, from vector projection, we observe that only the combined azimuthal and radial magnetic field can break the symmetry and cause a force imbalance leading to motion. 
This linear analysis indicates that perturbations of the droplet interface can propagate (and, since $v_p=v_p(k)$, they also experience dispersion). Such wavepackets will either decay or blow-up exponentially according to the sign of $\lambda(k)$. However, this is not the whole story, and \emph{nonlinearly stable} traveling shapes exist, as we now show.

\section{Nonlinear regime}
To demonstrate the possibility of nonlinear traveling waves in this system, we numerically solve the weakly nonlinear mode-coupling equations~\eqref{eq:wnl} for five modes (\textit{i.e.}, $k, 2k, \hdots, 5k$). The fundamental mode $k=7$ is chosen to allow propagating solutions over a wider swath of the $(\nba,\nbr,k_m)$ space (compared to choosing $k<7$), while only requiring modest spatial resolution for simulations (compared to $k>7$). We verified that the amplitudes $|c_n|$ and phases $\angle{[c_n]}$ of modes saturate at late times, leading to permanent propagating profiles with $\xi_{nk}(t)=c_n e^{in\omega(k) t}$ (see below). 

Next, we perform fully nonlinear simulations to validate the weakly nonlinear predictions. The vortex sheet method is a standard sharp-interface technique for simulating dynamics of Hele-Shaw flows \cite{R83}. It is based on a boundary integral formulation in which the interface is formally replaced by a generalized vortex sheet \cite{Prosp02} with a distribution of vortex strengths $\gamma(s,t)$, where $s$ is the arclength coordinate. We adapt this approach to handle ferrofluids under imposed magnetic fields. First, we express the velocity of the interface solely in terms of the interface position. To do so, it is convenient to identify the position vector in $\mathbb{R}^2$ with a scalar $z(s,t)\in\mathbb{C}$ ($^*$ denotes complex conjugate) \cite{Prosp02,S92,STW97}. Second, to advance the interface, we solve the dimensionless equations
\begin{subequations}\begin{align}
z_t^*&=-\frac{\gamma}{2z_s}
+\frac{1}{2\pi i}\mathcal{P}\oint\frac{\gamma(s',t)}{z(s,t)-z(s',t)}ds', \displaybreak[3]\\
\frac{\gamma}{2}&=\Re\left\lbrace
\frac{z_s}{2\pi i}\mathcal{P}\oint\frac{\gamma(s',t)}{z(s,t)-z(s',t)}ds'\right\rbrace
\nonumber\\
&\qquad
+\left[\kappa(s,t)-(\bbm M\cdot \un)^2
-\Psi\right]_s,
\label{eq:gamma}
\end{align}\label{eqs:vortex}\end{subequations}
iteratively for the velocity $z_t$, where $( \cdot )_t \equiv \partial(\cdot )/\partial t$, $( \cdot )_s \equiv \partial(\cdot )/\partial s$, $i=\sqrt{-1}$, and $\mathcal{P}$ represents principal value integration. Here, $\Psi=\nba r^{-2}+\nbr r^2$, and $(\bbm M\cdot \un)^2=\chi \left[\sqrt{\nba}r^{-1}(\et \cdot \un)+ \sqrt{\nbr}r(\er \cdot \un)\right]^2 $
is the dimensionless magnetic normal stress.
Time advancement is accomplished by the Crank--Nicolson scheme. The spatial discretization is implemented on an array of Lagrangian points ($N=1024$) with uniform $\Delta s$; see Appendix~\ref{sec:numerics} for further details, including algorithm flowchart and grid convergence study.

\section{Evolutionary dynamics}
\begin{figure*}
 \centering
 \includegraphics[keepaspectratio=true,width=\textwidth]{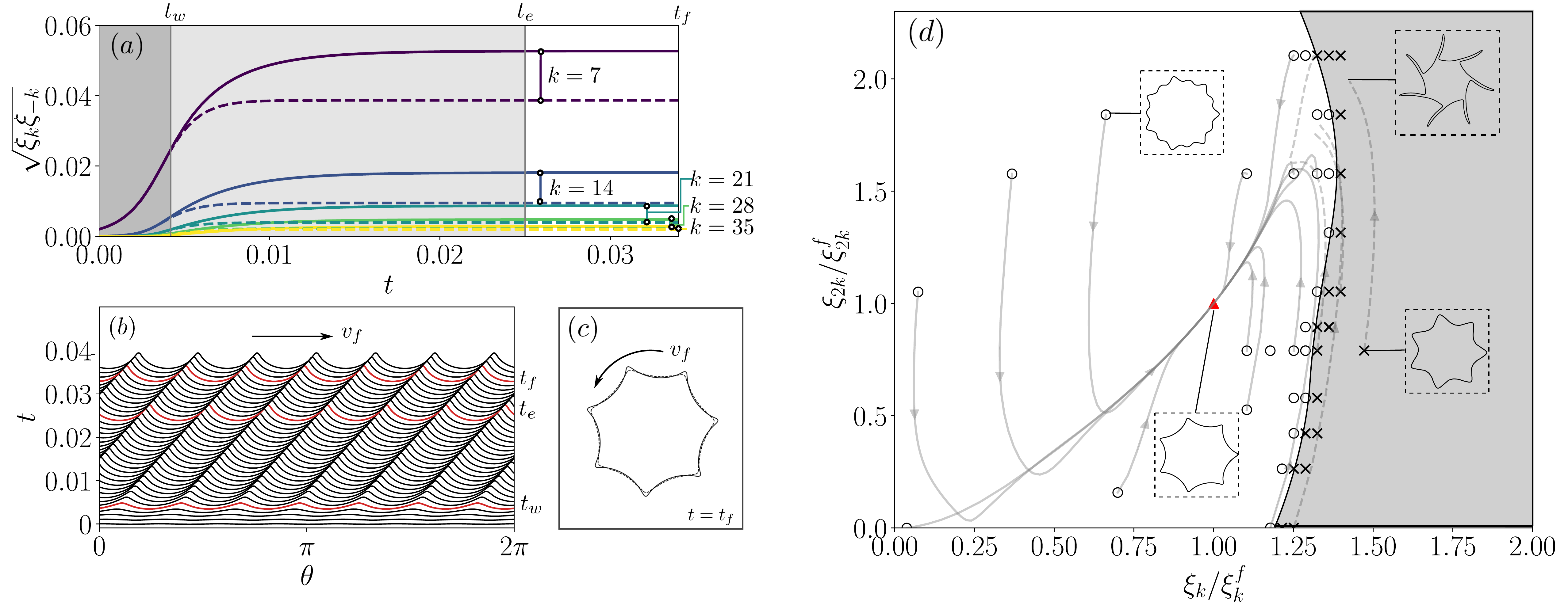}
 \caption{(a) The evolution of the first 5 harmonic modes from fully nonlinear simulation (solid) and weakly nonlinear approximation (dashed), for $\nba=1.0$, $\nbr=37$, $R=1$, and $\chi=1$ (same parameters for (b), (c) and (d)). (b) The fully nonlinear evolution of the interface from a small perturbation of the flat base state into a permanent traveling wave (rotating droplet). 
 (c) Comparison between the final shape from fully nonlinear simulation (solid) and weakly nonlinear approximation (dashed);
 (d) Stability diagram based on the first two harmonic modes of the final shape (marked with $\textcolor{red}{\blacktriangle}$) shown in (b); $\circ$ (resp.\ $\times$) denotes the stable (resp.\ unstable) initial conditions, solid  (resp.\ dashed) curves tract the stable (resp.\ unstable) evolution trajectories. The unstable region is shaded, and the `$f$' superscript represents the final harmonic mode amplitude.}
\label{fig:attractor}
\end{figure*}

The evolution of perturbed harmonic modes $\xi_k$ under the fully nonlinear simulation and the weakly nonlinear approximation are shown in Fig.~\ref{fig:attractor}(a). Starting from small initial values ($\xi_k|_{k=7}=0.002$, $\xi_{nk}=0$ for $n>1$) with $\nba,\nbr, R,\chi$ set so that the most unstable mode is equal to the fundamental mode ($k_m=k=7$), they saturate at late times. The perturbed circular interface grows exponentially in the linear regime and then matches the weakly nonlinear approximation at intermediate times ($t\in[0, t_w]$). The nonlinear simulations take longer to saturate ($t\in[t_w, t_e]$) and do so at higher final amplitudes compared to the weakly nonlinear result. The time-domain evolution is also shown in Fig.~\ref{fig:attractor}(b), evolving from a nearly flat (unwound circular) interface into a permanent propagating profile \footnote{See Supplemental Material at [URL will be inserted by publisher] for a video of this process.}.

The rotating droplet, shown in Fig.~\ref{fig:attractor}(c), has a polygonal shape with the symmetry set by the fundamental mode, $k = 7$. The fully nonlinear profile has a sharper peak compared to the weakly nonlinear approximation, which is otherwise in good agreement. The key discovery of the present work is the \emph{stable rotating shape}, which we now seek to analyze as a nonlinear wave phenomenon \cite{Re99}.

\section{When does weakly nonlinear stability imply nonlinear stability?}
A deficiency of linear and weakly nonlinear analyses is that they do \emph{not} provide sufficient conditions for stability. Linearly stable base states can be nonlinearly unstable \cite{KDS15}, and \textit{vice versa}. Importantly, however, our nonlinear traveling wave solution is a local \emph{attractor} (following the terminology from \cite{LLFP09}); see Fig.~\ref{fig:attractor}(d). 

Shapes in a neighborhood of the propagating profile, subject to small ($\xi_{k,2k}/\xi_{k,2k}^f \ll1$) or intermediate ($\xi_{k,2k}/\xi_{k,2k}^f = \mathcal{O}(1)$) initial perturbations, converge to it. Larger perturbations (shaded region) lead to \emph{nonlinear} instability of the weakly nonlinearly stable profiles; ``fingers'' continue to rotate and grow without bound under the effect of the radial magnetic field $\propto\nbr$, which increases with distance to the center of the droplet. 
Convergence to the attractor is sensitive to the initial amplitude of the first harmonic mode $\xi_k$. For the chosen parameters, $\lambda(k)>0$ and $\lambda (2k)<0$: high wavenumbers decay and the fundamental wavenumber grow in the linear stage. Consequently, for low $\xi_k/\xi_k^f$ and high $\xi_{2k}/\xi_{2k}^f$, the low wavenumber modes grow and saturate, as high wavenumber modes decay exponentially in the linear regime. With higher initial $\xi_k/\xi_k^f$, the  perturbed droplet will not go through the linear regime, and the amplitudes of both modes will rapidly increase to create a skewed shape, with multivalued $h(\theta,t)$, for which harmonic modes can no longer be defined. Note that Fig.~\ref{fig:attractor}(d) is a projection in the $(\xi_k,\xi_{2k})$ plane, where the initial values of $\xi_{3k}$, $\xi_{4k}$, $\xi_{5k}$ are set as the final amplitudes (and phases) from the weakly nonlinear equations. A fast Fourier transform was used to decompose the nonlinear profile into normal modes that we plot in this figure. Note that even though Fig.~\ref{fig:attractor}(a) indicates $\xi_{3k}$ makes a non-trivial contribution to the final shape, while $\xi_{4k}, \xi_{5k}$ play a smaller role, the projection is sufficient to conclude that the propagating wave profile is an attractor.

\section{Propagation velocity}
A permanent traveling wave profile has $\xi(\theta,t)=\Xi(k\theta-\omega t)$, and $v_f=\omega/k$ is its propagation velocity. Expressing the modes' complex amplitudes as $\xi_{nk}(t)=c_ne^{-in\omega(k) t}$, with constant $c_n\in\mathbb{C}$  that account for their relative phases, we have $v_p(k,t)=n\omega(k)/nk = v_f$. The mean $v_p$ of the first five harmonics is used to calculate $v_f^F$ for the fully nonlinear simulation and also $v_f^W$ for the weakly nonlinear approximation. Meanwhile, $v_f^L=v_p$ as given by Eq.~\eqref{eq:linear_vp}.

For a quantitative comparison, three sets of parameters are considered, fixing $\chi=1$. Two sets (i) and (ii) are for $k_m=7$, and the variation of $\nbr$ is according to Eq.~\eqref{eq:km}. A third set (iii) explores the effect of $k_m$ under the same linear propagation velocity $v_f^L$. Figure~\ref{fig:velocity}(a) compares the final propagating velocity predictions. Both $v_f^L$ and $v_f^W$ are in relatively good agreement with $v_f^F$ for small velocities. When $\nba\to0$ (the magnetic field becomes radial), only a stationary (non-rotating) droplet ($v_f=0$) exists~\cite{ALM18}. For higher $v_f$, the larger deviation in the predictions highlights the importance of nonlinearity. Nevertheless, the linear and weakly nonlinear results follow a similar trend. Importantly, $v_f^L$ and $v_f^W$ help identify the key control factors: the coupled magnetic field strength $\sqrt{\nba\nbr}$ and the radius of the initial droplet $R$. The salient physics uncovered is that the propagating velocity can be non-invasively tuned.

\section{Traveling wave shape}
\begin{figure}
 \centering
 \includegraphics[keepaspectratio=true,width=\columnwidth]{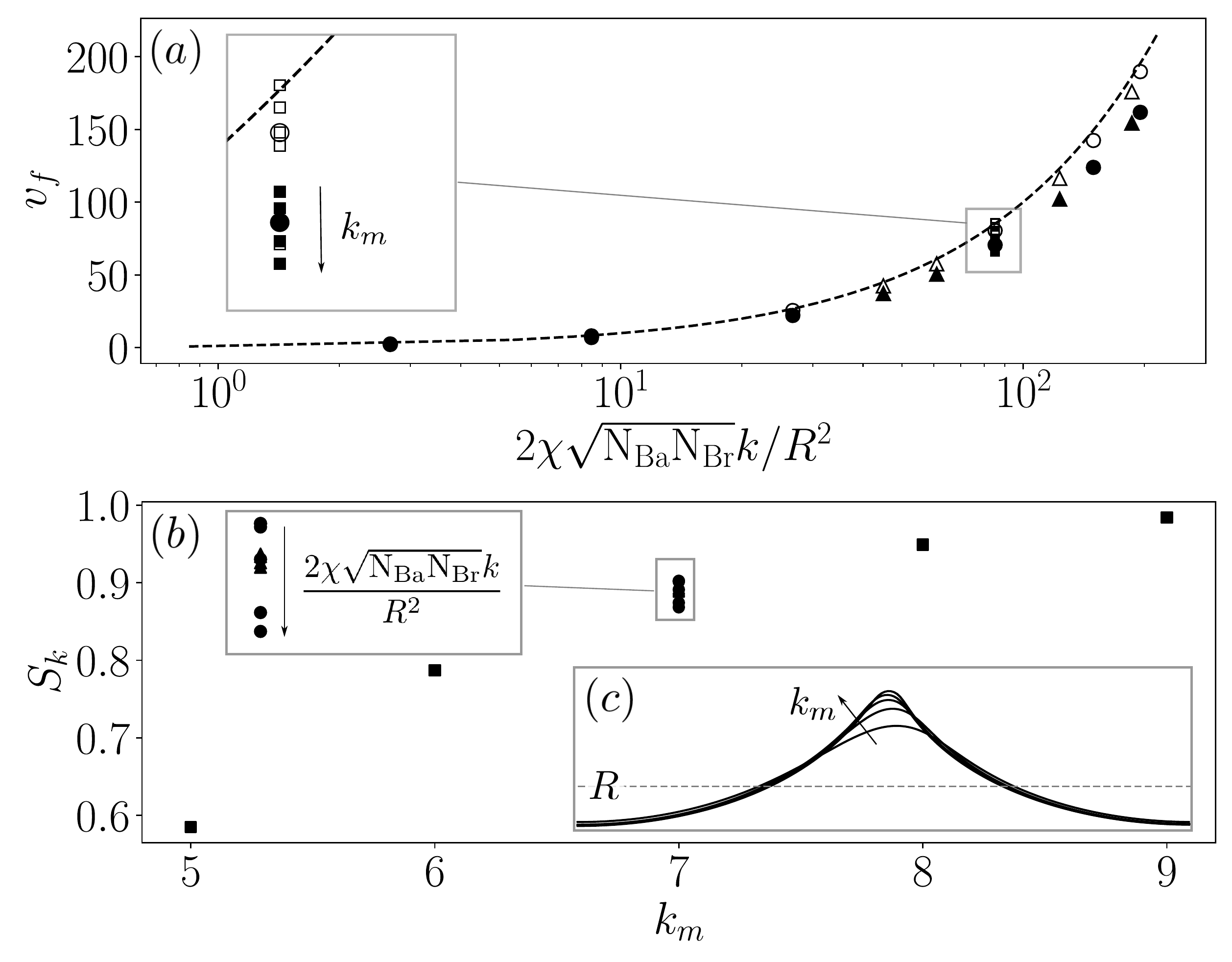}
 \caption{  
 (a) Comparison of the propagation velocity  predicted by linear theory $v_f^L$ (dashed), weakly nonlinear theory $v_f^W$ (empty symbols), and fully nonlinear simulation $v_f^F$ (filled symbols). The circles represent results for case (i) $R=1$ fixed and $\nba \in [0, 10^{-3}, 10^{-2}, 10^{-1}, 1, 3, 5]$, the triangles represent results for case (ii) $\nba=1$ fixed with $\nbr$ varying according to $R \in [0.8, 0.9, 1.1, 1.2]$, and the squares represent case (iii) $k_m \in [5, 6, 7, 8, 9]$, $R=1$ and $\nba,\nbr$ determined so that $v_f^L=85.16$.
 (b) The skewness $\sk$ of the fully nonlinear profile. 
 (c) The permanent wave shape (only one wavelength shown).
}
 \label{fig:velocity}
\end{figure}

The most unstable mode $k_m$ sets the propagating profile, which has a sharper peak for higher $k_m$ [Fig.~\ref{fig:velocity}(c)]. To quantify the shape change, we introduce the skewness $\sk(t) = \langle\xi^3\rangle/\langle\xi^2\rangle^{3/2}$, which is used to define the vertical asymmetry of nonlinear surface water waves~\cite{KCKD00,M13}; $\sk>0$  corresponds to narrow crests and flat troughs. Here, $\left\langle\, \cdot\, \right\rangle=\frac{1}{2} \pi\int_0^{2\pi} (\,\cdot\,)\, d\theta$. Figure~\ref{fig:velocity}(b) shows that $\sk$ (for the fully nonlinear propagation) increases with $k_m$, as expected from the sharper peaks in Fig.~\ref{fig:velocity}(c). This observation also explains why  $v_f^L$ becomes a worse approximation of $v_f^F$ as $k_m$ increases [inset of Fig.~\ref{fig:velocity}(a)]: smoother peaks (lower $k_m$) are better captured by the linear theory based on harmonic modes. 

Figure~\ref{fig:velocity}(c) reveals that the wave profile for $k_m=5$ ($\nba=1.9$, $\nbr=19.5$) is more asymmetric than the one for $k_m=9$ ($\nba=0.60$, $\nbr=60.8$). Under a purely radial magnetic field ($\nba=0$), the stationary shape has azimuthal symmetry~\cite{ALM18}. For the combined magnetic field, on the other hand, the dimensionless governing Eq.~\eqref{eq:governing} and pressure boundary condition in Eq.~\eqref{eq:pressure jump} can be rewritten as 
\begin{align}
\bbm v &=-\bbm\nabla\left(p-\nba \frac{1}{r^2}-\nbr r^2\right),\label{eq:non-governing}\\
p&= \kappa - \left[ \chi \frac{\nba}{r^2}(\et \cdot \un)^2
+\chi \nbr r^2(\er \cdot \un)^2 \right. \nonumber\\
&\qquad \left. +2\chi\sqrt{\nba \nbr}(\et \cdot \un)(\er \cdot \un)
\right],
\label{eq:non-PBC}
\end{align}
where $\et \cdot \un=-h_\theta/(h^2+h_\theta^2)$, $\er \cdot \un=h/(h^2+h_\theta^2)$, and $h_\theta=\partial h/\partial\theta$. The magnetic scalar potential in Eq.~\eqref{eq:non-governing} results from the body force, and the terms pre-multiplied by $\chi$ in Eq.~\eqref{eq:non-PBC} represent the magnetic normal stress. For a droplet with symmetric azimuthal perturbation, the body force alone cannot break the symmetry. Therefore, the asymmetry of shapes discussed is to be attributed to the magnetic normal stress. 

This observation can be intuitively understood by considering one wavelength of a symmetric waveform. The first three terms on the right-hand side of Eq.~\eqref{eq:non-PBC} are equal on both sides of the peak, while the fourth term changes at the peak due to the sign of $h_\theta$, which requires different curvatures on either side of the peak to remain balanced. Therefore,  $\sqrt{\nba\nbr}$ can be taken as the measure of the coupling effect between the magnetic field components.

\begin{figure}
 \centering
 \includegraphics[keepaspectratio=true,width=\columnwidth]{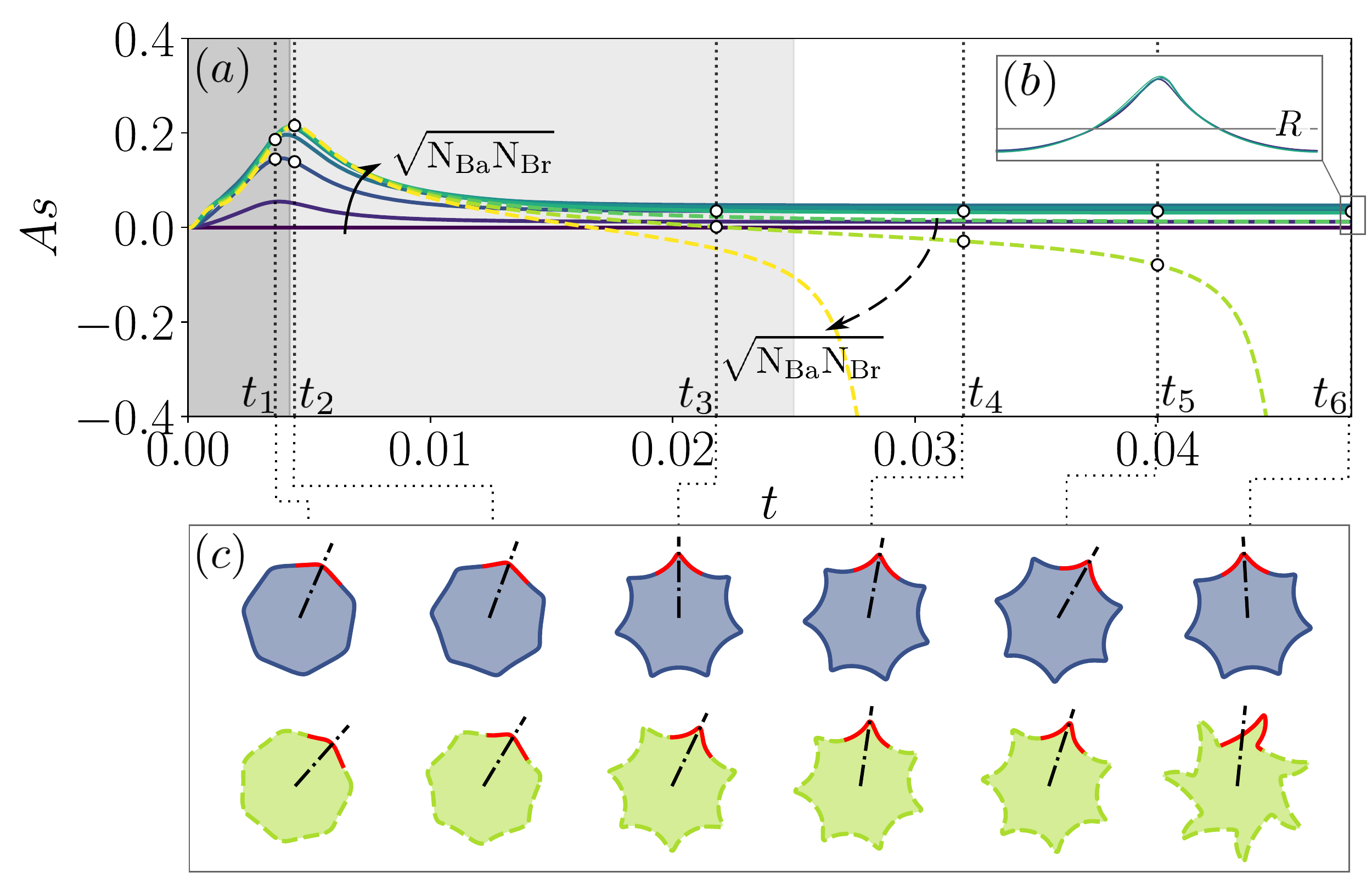}
 \caption{(a) Time evolution of the wave profile asymmetry for different combination of $\nba\in [0, 0.1, 1, 3, 5, 6, 7, 8, 9]$ and $\nbr$ varying so that $k_m=7$ for $R=1$. Solid curves represent stable cases yielding a  propagating profile; dashed curves represent unstable cases in which the profile distorts and grows without bound. (b) Permanent wave profiles that emerge and propagating in a stable manner. (c) Stable (top, with $\nba=1,\nbr=37$) and unstable (bottom, $\nba=8,\nbr=41$) evolution of the profile. The instants of time (at which the shapes in (c) are shown) are marked with white dots in (a), superimposed on the asymmetry profiles.}
 \label{fig:asymmetry}
\end{figure}

To further understand the asymmetry of propagating shapes induced by the combined magnetic field, we extend the parameters of case (i)  to a new case (iv): $k_m=7$, $R=1$ and $\nbr$ varying according to $\nba$ (see Fig.~\ref{fig:asymmetry} caption). To quantify the fore-aft asymmetry of the shape, we introduce  $\as(t) = \langle \mathcal{H}[\xi]^3\rangle/\langle\xi^2\rangle^{3/2}$ \cite{KCKD00,M13}; $\mathcal{H}[\,\cdot\,]$ is the Hilbert transform. For $\as>0$, waves tilt ``forward'' (\textit{i.e.}, counter-clockwise).

Figure~\ref{fig:asymmetry}(a) shows $\as(t)$ for different $\sqrt{\nba\nbr}$, which quantifies the coupled field effect, starting with small symmetric perturbations. For a stable case, $\as(t)$ reaches a maximum value ($t\approx t_1$) during the initial unstable weakly nonlinear growth (dark shadow region), and asymptotes to a value close to zero ($t\ge t_6$). The differences in the final propagating profile (under the same $k_m$) shown in Fig.~\ref{fig:asymmetry}(b) are hard to capture, which is consistent with the observation in Fig.~\ref{fig:velocity}(b).
For the unstable cases, ``wave breaking'' occurs, which is highlighted by a change of sign of $\as$. Also, now, $\as(t)$ no longer saturates at late $t$. Instead $\as(t)$ crosses zero (at $t\gtrsim t_3$) and approaches a singularity. This unstable example is shown in the second row of Fig.~\ref{fig:asymmetry}(c). As its amplitude first grows, the wave tilts forward ($t=t_2$), but nonlinear effects restore its symmetry ($t=t_3$). Subsequently, the wave tilts backwards ($t=t_4,t_5$) and its amplitude continues to grow ($t=t_6$). The calculation of $\as$ then fails because $\mathcal{H}$ requires the perturbation $\xi(\theta,t)$ to be single-valued in $\theta$. The distorted wave has a wider base and evolves into long unstable fingers.

Note that $\nba/\nbr$ also increases with $\sqrt{\nba\nbr}$ for our choices of $\nba$ and $\nbr$. Equation~\eqref{eq:Linear growth} shows that the radial magnetic field is destabilizing, while surface tension ($k>2$ here) and the azimuthal field are stabilizing. However, the nonlinear simulations indicate that, for the same $k_m$, increasing $\nba/\nbr$ can induce instability because it engenders a larger $v_f$ (and $\as$), leading to a global bifurcation with Fig.~\ref{fig:attractor}(d) as one stable slice. This result has an analogy to solitary waves in equations of the Kortweg--de Vries (KdV) type. Specifically, initial perturbations grow, deforming a shape until nonlinearity is balanced by dispersion, when a permanent wave emerges \cite{ZK65}. However, depending on the form of the nonlinearity, not all such permanent waves are stable attractors, and conditions must be placed on the wave speed \cite{BSS87}. 

\section{Conclusion} 
This study demonstrates how a perturbed circular ferrofluid droplet can evolve into a nonlinearly stable rotating shape. The most unstable mode sets how perturbations evolve into a permanent profile (and its skewness and asymmetry). Weakly nonlinear theory, in hand with fully nonlinear simulations, revealed permanent rotating shapes (traveling waves) with predictable propagation velocity. We showed how the coupling of the magnetic field components modifies the asymmetry and the nonlinear instability.

Although the manipulation of the linear growth rate of interfacial perturbations in Hele-Shaw cells is well studied \cite{MMM19}, including extensions based on the weakly nonlinear expansion from Eq.~\eqref{eq:wnl} \cite{DM2010}, the \emph{control} of the dynamic, \emph{fully nonlinear}, patterns is not. Our approach harnesses the magnitude and the direction of coupled magnetic fields to generate ferrofluid droplets, with well-characterized shapes and rotational speeds, by purely \emph{external} means.

Open questions remain: \textit{e.g.}, which fundamental modes evolve into propagating shapes? Work on the stationary problem~\cite{LM16,LOM08} gives a hint, however, for a propagating shape the Birkhoff integral equation~\cite{Birkhoff54} must be solved, making an extension of \cite{LM16,LOM08} challenging. Interestingly, our simulations also reveal that patterns predicted as stable by weakly nonlinear analysis can be unstable. In Appendix~\ref{sec:unstable_k4}, we provide an example showing that perturbations with $k=4$ will not evolve into either a stationary or a propagating shape (although both are predicted to exist by weakly nonlinear analysis). 

Additionally, does this system  accommodate more than one propagating wave? If so, do such waves keep their shapes upon collision, as with soliton interactions \cite{ZK65,Soomere2009}? Previous studies derived KdV equations for unidirectional small-amplitude, long-wavelength disturbances on fluid--fluid interfaces in Hele-Shaw \cite{ZY91} and axisymmetric ferrofluid configurations \cite{RE06,BBF10}, demonstrating the celebrated ``$\sech^2$'' solitary wave. Instead, in our study without such restrictions, we discovered \emph{periodic} traveling nonlinear waves, which are akin to the \emph{cnoidal} solutions of periodic KdV, \textit{i.e.}, the fundamental nonlinear modes (``soliton basis states'') \cite{OSBC91}. Additionally, we observed wave breaking [Fig.~\ref{fig:asymmetry}(c,bottom)]. 

Finally, it would be of interest to verify the proposed shape manipulation strategies by laboratory experiments. Previous theoretical studies~\cite{LM16,LMO10,DLM05,AOC04} suggest that many exact stationary droplet shapes are unstable, thus their relevance to experimental studies is limited. On the other hand, the nonlinear simulations in our study, showing stable rotation, pave the way for future experimental realizations. 

\begin{acknowledgements} 
This material is based upon work supported by the National Science Foundation under Grant No.~CMMI-2029540.
\end{acknowledgements}

%

\clearpage
\appendix

\section{Unstable pattern with $k=4$}
\label{sec:unstable_k4}

\begin{figure}[H]
 \centering
 \includegraphics[keepaspectratio=true,width=\columnwidth]{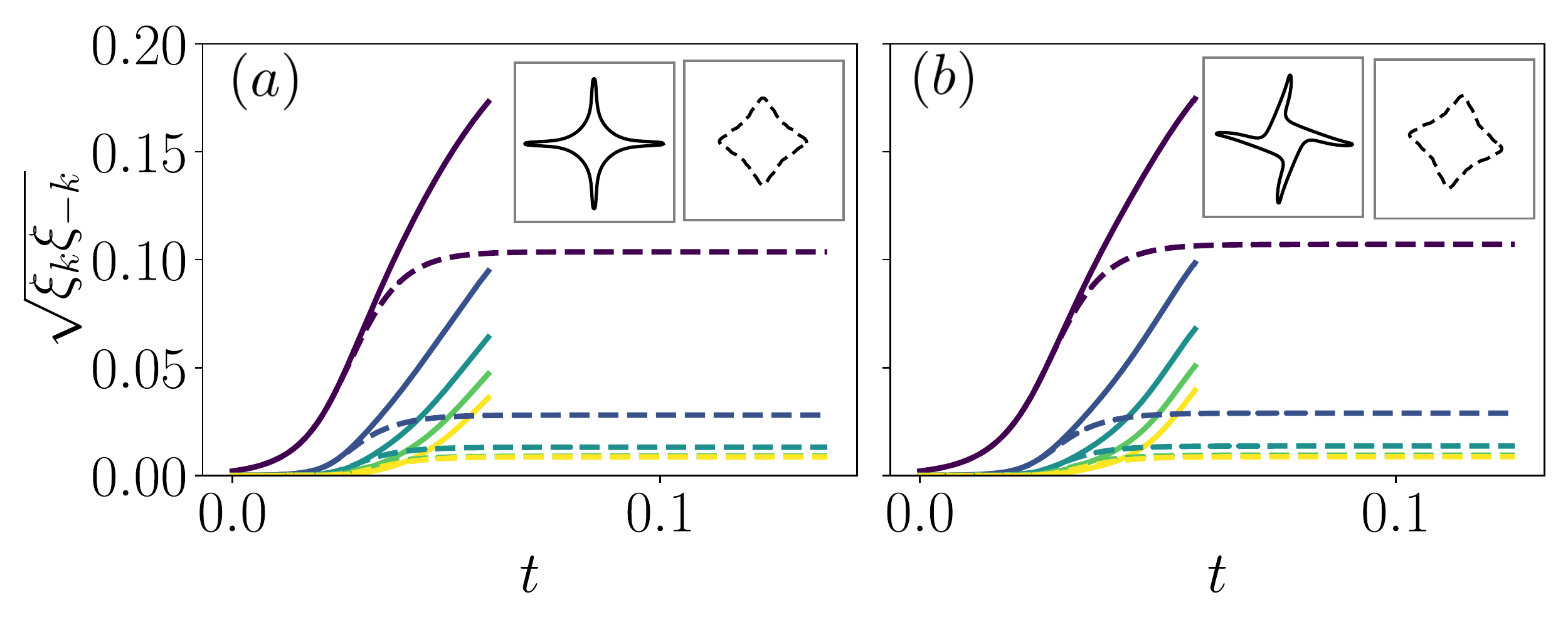}
 \caption{The unstable evolution of the first 5 harmonic modes ($k=4,8,12,16,20$) from fully nonlinear simulation (solid) and their stable evolution from weakly nonlinear approximation (dashed) for (a) nonrotating ($k_m=4$, $\nba=0$) and (b) rotating ($k_m=4$, $\nba=1$) shapes.
 }
 \label{fig:unstable k4}
\end{figure}

As mentioned in the main text, linear (or even weakly nonlinear) stability does not always imply nonlinear stability (see also \cite{KDS15}). Indeed, it is not known under what conditions the weakly-nonlinear stable droplet shapes are actually nonlinearly stable. In the main text, we presented examples for which this implication holds true. Here, in Fig.~\ref{fig:unstable k4}, we demonstrate, for completeness, an example to the contrary. To the best of our knowledge, such an example has not been analyzed before, and thus remains an avenue of future work. This exploration also requires caution, to rule out physical from numerical instability.

\section{Implementation of the numerical method and grid convergence}
\label{sec:numerics}

\begin{figure*}
    \centering
    \includegraphics[keepaspectratio=true,width=0.6\textwidth]{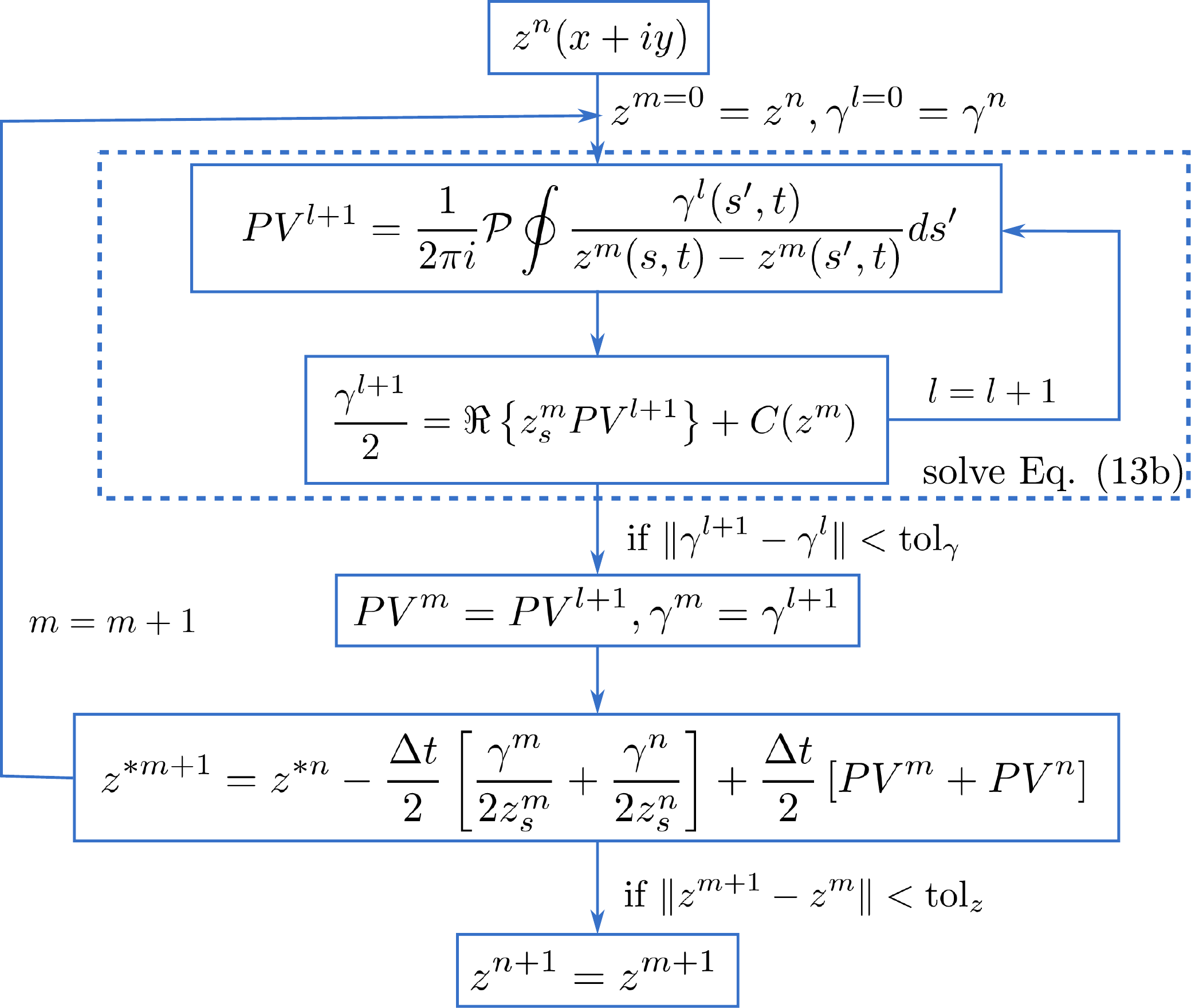}
    \caption{Flow chart of the vortex-sheet algorithm using Crank--Nicolson for time advancement and fixed-point iteration for resolving the implicit nonlinear terms.}
    \label{fig:flowchart}
\end{figure*}

The principal value integration in Eqs.~\eqref{eqs:vortex} is performed numerically by a spectrally accurate spatial scheme~\cite{S92}:
\begin{equation}
   PV_{j}=\frac{2 \Delta s}{2\pi i}\sum_{j+k \; \text{odd}}\frac{\gamma_{k}}{z_j-z_k},
\label{eq:spac discre}
\end{equation}
where a $j$ subscript denotes the evaluation of a quantity at the $j$th Lagrangian grid point $j\Delta s$ with $\Delta s=L/N$, $L=\oint ds$, and $N$ is the number of grid points. The parametrization of the interface via its arclength reduces the stiffness of the numerical problem caused by the presence of third-order spatial derivatives. A rearrangement of the grid points is conducted with cubic interpolation, after each time step, to maintain uniform grid spacing $\Delta s$. The uniform arclength spacing then allows the use of the second-order central differentiation formul\ae\ for all derivatives. A fixed-point iteration scheme is used to resolve the implicit Eq.~\eqref{eq:gamma} to obtain $\gamma_j$ at each interface point $z_j$, as shown schematically in Fig.~\ref{fig:flowchart}.
 
Time advancement (superscripts denote the time step number) is accomplished with a Crank--Nicolson scheme:
\begin{equation}
\begin{aligned}
    z^{*n+1}=z^{*n}
&-\frac{\Delta t}{2}\left[
\frac{\gamma^{n+1}}{2z_s^{n+1}}+
\frac{\gamma^n}{2z_s^n}
\right]\\
&+\frac{\Delta t}{2}\left[
PV^{n+1}+PV^n
\right],
\end{aligned}
\label{eq:CN}
\end{equation}
where both of the nonlinear terms $\gamma^{n+1}/2z_s^{n+1}$ and $PV^{n+1}$ are obtained by subiteration with index $m$, as shown schematically in Fig.~\ref{fig:flowchart}.  Equation~\eqref{eq:CN} converges and $z^{n+1}=z^{m+1}$ when $\| z^{m+1}-z^{m}\| <\text{tol}_z$ with $\text{tol}_z= 0.1\max|z_t^n|\Delta t$.

\begin{figure*}
   \centering
   \includegraphics[keepaspectratio=true,width=\textwidth]{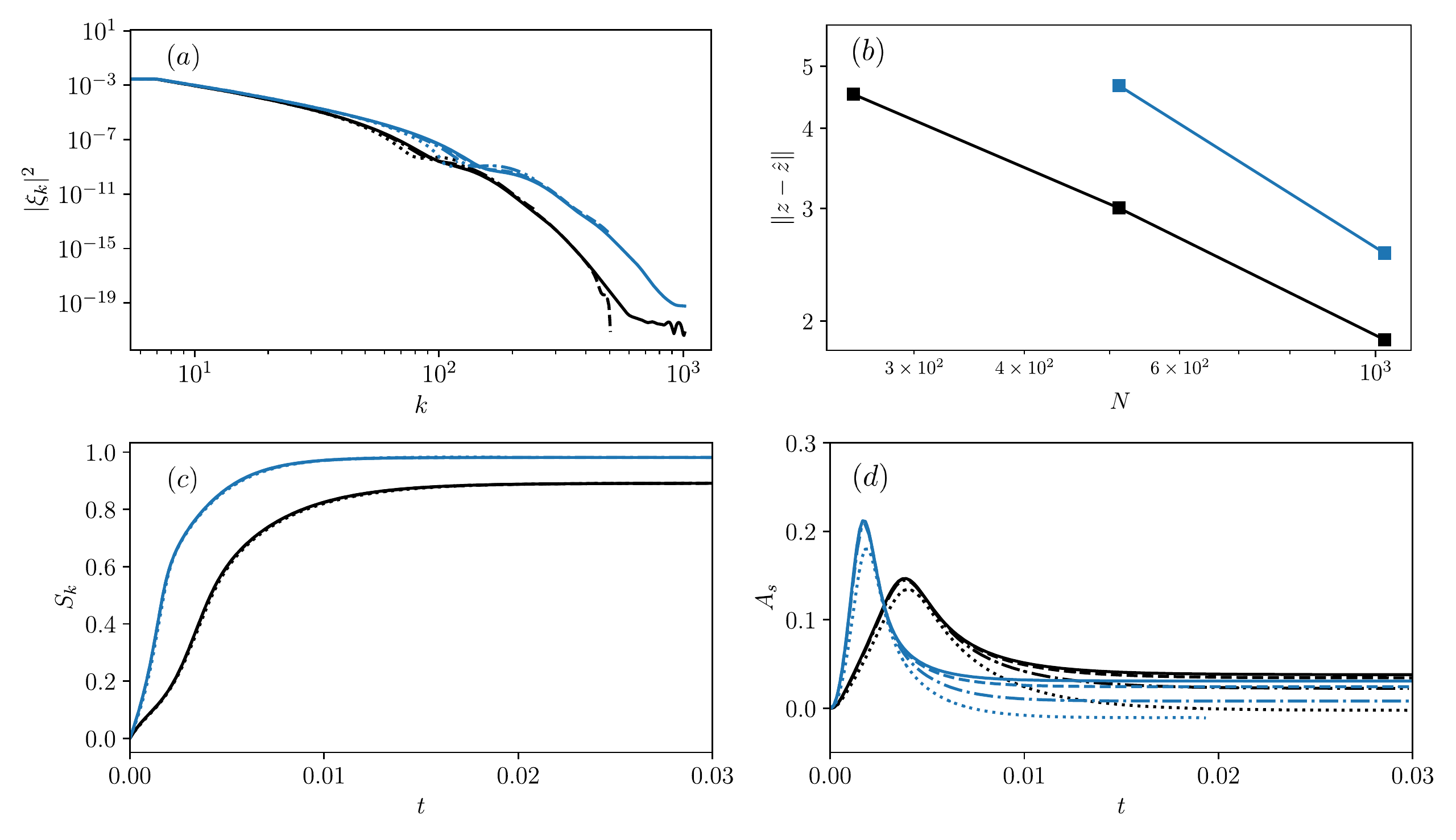}
   \caption{Grid convergence study for fundamental modes $k_m=7$ (black) and $k_m=9$ (blue) with $\nba=1$ and $N=256$ (dotted), $N=512$ (dot-dashed), $N=1024$ (dashed), and $N=2048$ (solid). (a) Spectral energy of harmonic modes ($k$, $2k$, $3k$, $\hdots$). (b) The root-mean-square error taking the $N=2048$ solution $\hat z$ as ``exact''. (c) Grid convergence of the evolution of the skewness $\sk(t)$. (d) Grid convergence of the evolution of the asymmetry $\as(t)$.}
   \label{fig:grid_conver} 
\end{figure*}

A grid convergence study with 4 levels of the grid resolution was conducted for two cases: $k_m=7$ and $k_m=9$ with $\nba=1$. The most frequently used case in the main text is $k_m=7$, while the sharper peaks for $k_m=9$ demand on the highest grid resolution.
Figure~\ref{fig:grid_conver}(a) shows the spectral energy content of harmonic modes ($k$, $2k$, $3k$, $\hdots$) of the propagating waveform, where the ``piling up'' near the tail on the finest grid ($N=2048$) is numerical noise. This plot supports our decision to consider the $N=1024$ grid as offering sufficient resolution. Figure~\ref{fig:grid_conver}(b) shows the root-mean-squared error in the shape $z$ itself, taking $\hat z$ as the ``reference shape'' on the $N=2048$ grid. The error decreases with grid refinement. The error at $N=256$ for $k_m=9$ is not shown for the propagating shape because the scheme is not even stable on such a coarse mesh for this case.

Figure~\ref{fig:grid_conver} further shows the evolution of (c) the skewness $\sk$ and (d) the asymmetry $\as$. The skewness matches well on all grids used, showing it is a well-converged quantity, while the asymmetry is seen to be more sensitive to the grid resolution. The differences between $N=1024$ and $N=2048$ are small enough so that it is safe to use $N=1024$ for the simulations reported in the main text, considering the significantly higher computational cost incurred by using finer grids.

\end{document}